# Main-path analysis and path-dependent transitions in *HistCite*™-based historiograms

Diana Lucio-Arias[1] & Loet Leydesdorff[2]
Amsterdam School of Communications Research (ASCoR), University of Amsterdam
Kloveniersburgwal 48, 1012 CX  Amsterdam, The Netherlands.
[1] tel.: +31-20- 5253753; fax: +31-20- 525 3681;  D.P.LucioArias@uva.nl
[2] loet@leydesdorff.net; http://www.leydesdorff.net

With the program *HistCite*™ it is possible to generate and visualize the most relevant papers in a set of documents retrieved from the *Science Citation Index*. Historical reconstructions of scientific developments can be represented chronologically as developments in networks of citation relations extracted from scientific literature. This study aims to go beyond the historical reconstruction of scientific knowledge, enriching the output of *HistCite*™ with algorithms from social network analysis and information theory. Using *main path analysis,* it is possible to highlight the structural backbone in the development of a scientific field. The expected information value of the message can be used to indicate whether change in the distribution (of citations) has occurred to such an extent that a *path-dependency* is generated. This provides us with a measure of evolutionary change between subsequent documents. The "forgetting and rewriting" of historically prior events at the research front can thus be indicated. These three methods are applied to a set of documents related to *fullerenes* and the fullerene-like structures of *nanotubes*.

**Introduction**

Derek de Solla Price (1965) proposed using the literary model as a functional simplification of the process of scientific discovery and communication. In this model the dynamics, developments, and structure of science are operationalized in terms of networks of scientific papers (Garfield, 1979). The model was originally applied to trace scientific developments historically (Garfield *et al*., 1964), but was also used to study scientific specialties (Griffith *et al*., 1974; Small & Griffith, 1974), influences in science (Stewart, 1983), etc. Given that publishing their scientific results is one of the scientists' main and perhaps most relevant activities, this operationalization may provide an accurate proxy for the study of scientific developments. Through the literary model, scientific developments can be associated with the accumulation of scientific accomplishments.

Citations can be considered as unidirectional links that relate later documents to earlier ones (Garfield, 1973; Small & Griffith, 1974). The historical dependency of scientific developments operates through citations and references so that citation patterns are associated with interpretation of previous results, successful papers (Small, 1978), and the intention of scientists to position their results differently from previous ones (Fujigaki 1998a). This citation culture (Wouters, 1999) can be used to understand scientific developments in terms of patterns of references emerging and reproduced in scientific literature.

Citations can be analyzed as links between authors and/or texts (Leydesdorff & Amsterdamska, 1990). Using the literary model, citation relations are considered as links among different texts holding cognitive significance. Citations made to other documents position the citing document with reference to papers in an evolving network. The position of each paper can be expected to change over time as the research front further develops and older citations become obsolete or are overwritten by newer ones (Fujigaki, 1998a and b; Garfield, 1963; Merton, 1979, at p. vii).

The citation relation that links two documents reveals two different dynamics in the process of scientific development: codification and diffusion. In the first part of this study, a citation will be considered as codification: a "citing" document makes reference to a body of knowledge that is further codified by this reference among other possible references (Leydesdorff & Wouters, 1999). In the second part, we invert the direction and citation will be considered as an "is cited by" relation. This reflects the diffusion of knowledge claims from an original document to documents published thereafter. While codification is a reflexive process taking place in the present and reconstructing the past, the diffusion of texts can only be measured by following the arrow of time in the forward direction.

This work builds on a previously published study which provided descriptive statistics and substantive analysis of documents retrieved from the ISI *Web of Science* with "fullerene*" (7,696 documents) or "nanotube*" ( 9,672 documents) among their title words (Lucio-Arias & Leydesdorff, 2007). Fullerenes and nanotubes were chosen as they can be considered the subjects of closely related scientific research fronts. Fullerenes were discovered as molecules in 1985; carbon nanotubes were discovered in 1991 as special compounds with fullerene-like structures. In other words, not every fullerene can be regarded as a carbon nanotube. The literatures on both topics were closely related in the early 1990s, but the respective systems of scholarly communication seem to have differentiated increasingly thereafter (Figure 1).

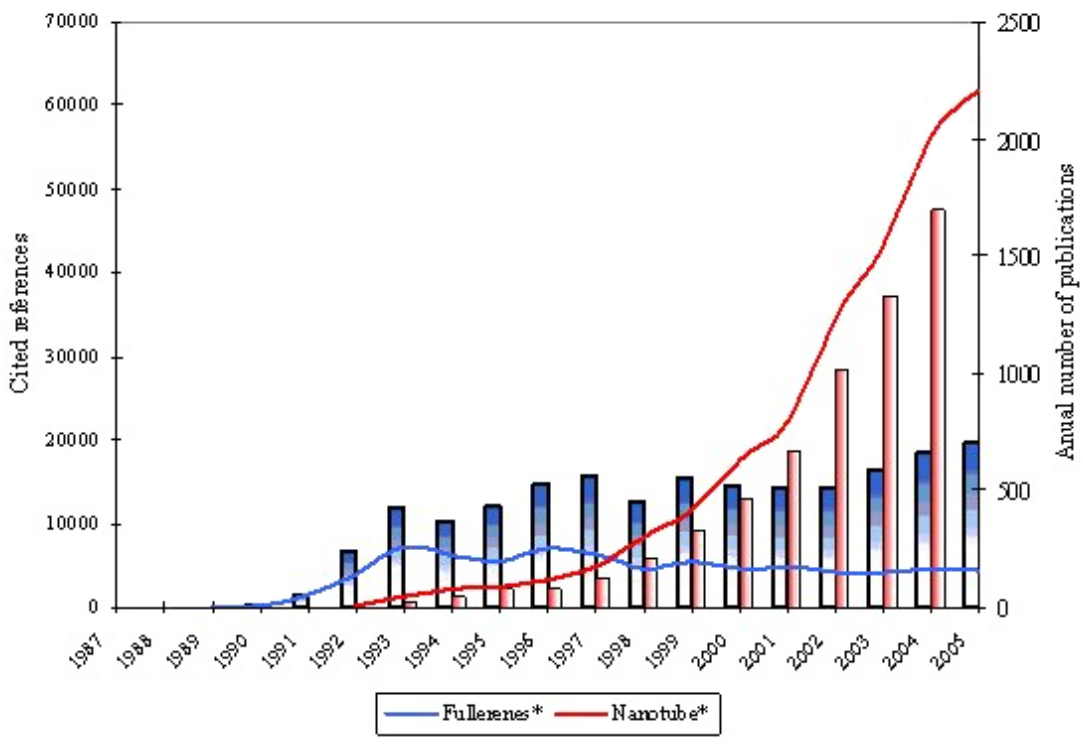

**Figure 1.** Documents in the SCI with "fullerene*" or "nanotube*" in their titles over time. The lines indicate the number of publications (legend on the right axis) and the bars the aggregated number of cited references per year (legend on the left axis).

In this paper, algorithmically built historiograms are used to extend the notion of two different specialties by analyzing the structures among the most frequently cited documents for each of the sets. The citing and cited relations will be analyzed in terms of codification and diffusion processes using network analysis and entropy measures. In other words, the focus of this paper is methodological: can algorithmic historiography further be extended by using quantitative measures?

For both fullerenes and nanotubes, a high aging rate of citations can be expected because these are newly emerging fields of science (Price, 1965; Chen, 2005). One can expect the short-time citation window to be reinforced by the fact that in a dynamic field of research there is a tendency to break important contributions into subsequent papers corresponding to the sequential stages of their development (Garfield *et al.*, 1964). However, the number of documents considered (7,696 and 9,672, respectively) should be enough for a citation network that represents the historical development of research in each of these fields.

The operationalization in terms of the literary model enables us to apply algorithms in order to visualize trends associated with the development of science (Burt, 1983). Hitherto, most of the visualization tools available to map scientific literature using citation and co-citation analysis represent the state of science at specific moments of time. *CiteSpace* (Chen, 2005) diachronically visualizes the evolution of knowledge domains and research fronts. *HistCite* makes it possible to map the historical evolution of a set of papers in terms of trajectories: it generates historiograms which reflect the flow of ideas up to the moment of the last document considered. Additionally, it provides output files that can be read by software for social network analysis.

Following this section, the use of *HistCite* will be explained along with a demonstration of how its output can be enhanced with measures obtained from social network analysis and information theory, notably main path analysis and the study of path-dependent transitions. Results for the cases of fullerenes and nanotubes follow the methods section. Addressing the citation relations as a flow of knowledge from one paper to the next will provide us with a perspective other than the hindsight perspective of the citing papers. This diffusion mechanism as different from the codification process is discussed in the fourth section.

**Methods**
    *a) HistCite*

The program *HistCite* creates a citation index for any defined set of documents retrieved from the ISI *Web of Science,* outlining the chronological network of citations among these documents (Garfield *et al*., 2002, 2003a, 2003b). The most significant works in a set are identified on the basis of citation frequencies, and the citation relations among these documents can be visualized. Because the tool highlights the most cited papers, their bibliographic antecedents and descendents both within and outside the set can be traced, facilitating the process of the historical reconstruction of fields of science. Garfield (2001) described this process as writing an "algorithmic historiography." The algorithmic approach enables us to include more variety in the perspective than does a historical reconstruction based on a single narrative (Kranakis & Leydesdorff, 1989).

*HistCite*'s algorithmic historiograms illustrate the key works associated with the development of a field, highlighting the most frequently cited papers in each collection. If the development of the sciences is understood as a series of chronological events (Garfield *et al*., 1964), and the citation network is acknowledged as an emergent property of the scientists' activities (Fujigaki, 1998a), one can consider that the network formed by relations between the most frequently cited documents represents the intellectual base from which further developments of the field unfold.

Using *HistCite*, the 30 most highly cited documents in each set of documents are identified. The citation relations between these documents can be exported to other programs for further analysis. For the analysis of main paths, we use Pajek, a freely available software program for the analysis and visualization of networks.[1] For the analysis of path-dependent transitions we wrote our own routines.[2] The algorithms used in this study enhance the representations from *HistCite* by qualifying the links with quantitative measures.

---

[1] Pajek is freely available for non-commercial use at http://vlado.fmf.uni-lj.si/pub/networks/pajek/
[2] Available at: http://home.medewerker.uva.nl/d.p.lucioarias/ or http://www.leydesdorff.net/software/crittrans/index.htm

Main-path analysis associates the links (citations) to their connectivity qualities; relative entropy (Kullback & Leibler, 1951) will be used as a measure to evaluate whether the distributions of references changed between papers to such an extent that a path-dependency was created (Mei & Zhai, 2005). The three measurement outputs can then be combined into the visualization of *HistCite*: they show, respectively, prominence and relevance, structural connectivity, and evolutionary dynamics. While relevance is measured by considering the number of citations a given document accumulates over time, main-path analysis considers both the citations which a document receives and the documents it cites. The analysis of path-dependent transitions focuses on the (distributions of) cited references. All these measures can be associated with the positional attributes of a paper in its network of relations (Burt, 1982).

### b) *Main-Path Analysis*

Main-path techniques examine connectivity in acyclic networks and are especially interesting when nodes are time dependent, as it selects the most representative nodes at different moments of time. In a citation network, time assigns direction to the links and each node represents a distinct event in time (Carley *et al.*, 1993).[3] A node that links many nodes and has many nodes linking to it will probably be part of the main path. The main path will highlight those papers that build on prior papers but continue to act as an authority in reference to later works (Yin *et al.*, 2006)

The main path is reconstructed by calculating the connectivity of the links in terms of their degree centrality and outlining the path formed by the nodes with the highest degree. In terms of a citation network, this degree measure considers the number of citations a document receives (indegree) as well as the number of cited references in the documents (outdegree). The main path is constructed by selecting those connected documents with the highest scores until an end document is reached (Batagelj, 2003). This can be either a document that is no longer cited or one that contains no further references within the set. *HistCite*'s output shows citations as a "cited-by" relation from the perspective of the citing documents. In the introduction, we distinguished this codification process from the diffusion process of knowledge claims in cited documents. In order to draw the main path of the most influential documents over time, it is necessary to transpose the matrix.

Citing previous literature and being cited by subsequent literature positions a paper in relation to other papers in the set (Hummon & Doreian, 1989). By constructing these positions, main-path algorithms enable us to make the structural backbone of a literature visible. When a set of documents represents a self-contained field—not significantly building on knowledge from other fields—the citation network among the key documents (the most highly cited ones) can be expected to contain at least one main path (Carley *et al.*, 1993).

---

[3] Main-path analysis cannot be applied to cyclic networks were nodes can belong to paths that lead back to themselves.

Three models to identify the most important part of a citation network can be distinguished: the *Node Pair Projection Count*, which accounts for the number of times each link is involved in connecting all node pairs; the *Search Path Link Count*, which accounts for the number of all possible search paths through the network emanating from an origin; and the *Search Path Node Pair*, which accounts for all connected vertex pairs along the paths (Hummon & Doreian, 1989, at pp. 50-51). Of these three methods, algorithms to estimate the latter two are included in Pajek (Batagelj, 2003).

In this study, we use the Search Path Link Count algorithm for the following reasons. The Search Path Node Pair chooses a path while forgoing citations between documents that are connected indirectly through a third document that may also be part of the path. For example, if there are citations between documents 1 and 2, documents 2 and 3, and documents 1 and 3, the main path calculated with the Search Path Node Pair will not consider the citation relation between document 1 and document 3. The Search Path Link Count is the preferred algorithm for this analysis because all citation relations are taken into account. This inclusive approach of all citation relations accords with the analysis of *HistCite* and path-dependent transitions.

### c) Path Dependency and Critical Transitions

By applying main-path analysis as described above, documents that summarize the main stream of research in fullerenes and nanotubes can be highlighted. These documents can be expected to contain the main findings in the fields, either because they are crucial in defining the field's cognitive history or because they represent high impact documents. In addition to connecting to other documents, however, each scientific text adds new information to the network. This information is contained in the distribution of attributes.

The expected information value $I$ of an additional text can be expressed as a Kullback-Leibler (1951) divergence between the *a prior* and *posteriori* distributions of attributes in each of the texts. The similarity between texts can thus be measured. This relative entropy measure is formalized as follows:

$$I(q:p) = \sum_{i=1}^{n} q_i \log_2 \left( \frac{q_i}{p_i} \right) \qquad (1)$$

In this equation, $p_i = (p_1, p_2, \ldots, p_n)$ represents the *a priori* distribution of references (in the first text) and $q_i = (q_1, q_2, \ldots, q_n)$ the *posterior* one (that is, in the next text considered). When two is used as the basis of the logarithm, $I$ is expressed in bits of information. $I$ is the expected information value of the message that the *a priori* distribution is transformed into the *a posteriori* one (Kullback & Leibler, 1951; Theil, 1972; Leydesdorff, 2005). Note that $I$ is asymmetrical in $p$ and $q$.

In each text, a number of distributions of attributes can be analyzed: words, title words, cited references, citations (Leydesdorff, 1995). Because we are interested in this study in processes of codification and diffusion, we transformed the reference lists of each text into a relative frequency distribution of occurrence (*f*) of each reference normalized at the level of the set ($f_i/N$). This results in size-equivalent vectors of distributions for each of the 30 most frequently cited documents that are used to measure change in the citation patterns among the documents. However, a zero in the *a priori* distribution (the denominator in Eq. 1) would make a non-zero value in the *posteriori* distribution a complete surprise, and the expected information content of the message that this happened would therefore be infinite. This would distort the analysis. For this reason, the values of the cells were increased in a unity for this analysis (Elliot, 1977; Price, 1981).

The Kullback-Leibler divergence can be used to analyze path-dependent transitions in a set of sequential events (Frenken & Leydesdorff, 2000; Leydesdorff, 1995, at p. 341). If the prediction of the distribution of references in the *a posteriori* text on the basis of the *a priori* one were perfect, the expected information content of the message that the new text arrived would be zero. The paper would be a copy of the previous one in terms of its cited references and nothing would have changed.

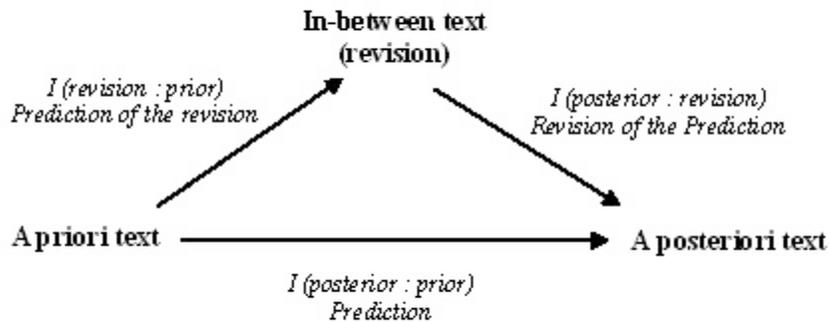

**Figure 2**. Prediction and possible revision of the prediction among three documents

If the prediction is imperfect, it can be improved by an in-between text (Figure 2). This improvement of the prediction of the *a posteriori* probability distribution ($\Sigma_i\, q_i$) on the basis of an in-between probability distribution ($\Sigma_i\, p_i'$) compared with the original prediction ($\Sigma_i\, p_i$) can be formulated as follows:

$$I(q:p) - I(q:p') = \sum_i q_i \log_2(q_i/p_i) - \sum_i q_i \log_2(q_i/p_i')$$

$$= \sum_i q_i \log_2(p_i'/p_i) \qquad (2)$$

These revisions of the prediction can occur among any three documents in a set. If $I(q:p) > I(q:p') + I(p':p)$ the pathway in Figure 2 via the revision is a more efficient channel for the communication between sender and receiver than their direct link. Contrary to the geometry of Figure 2, the sum of the information distances via the intermediate document has become shorter than the direct information path between the sender and the receiver.

Unlike the evolutionary transitions as defined by Mei & Zhai (2005, at p. 201), the identification of these critical transitions does not require the specification of a threshold because their identification is based on the solution of an inequality. On a timeline, these critical transitions indicate documents that can be considered as path-dependencies or obligatory passing points (Callon, 1986) in the sense that the later documents (the *a posteriori* text and the revision) contain similarities in the distribution of their cited references which the earlier ones did not have. The communication system has changed in the dimension of the distribution(s) under study.

In the case of diffusion, a critical transition is examined from the perspective of the sender: the intermediate document in this case boosts the signal from the *a priori* document as an auxiliary transmitter. The history before the intermediate station is overwritten. In the case of codification, the intermediate document provides a chronologically closer alternative to define the cognitive position of the cited document. Thus, it reinforces the codification of the latter's citation pattern in the archive because the newly added document no longer makes a difference for this position.

In general, information-theoretical measures allow for the extension to higher dimensionalities in the distribution, for example, by combining citations with title words, author names, institutional addresses, etc. The multivariate distributions remain fully decomposable (Theil, 1972). By extending the number of subscripts ($p_i$, $p_{ij}$, $p_{ijk}$, etc.), the measurement can further be *refined* (Van den Besselaar & Heimeriks, 2006). In this study, however, we focus on citations in order to show how these algorithms enable us to enrich the insights obtained from using *HistCite*.

The analysis of path-dependent transitions provides a meaning to the links that is different from the main-path analysis obtained with social network analysis. Main-path analysis identifies a continuous path of connected and connecting nodes, while critical transitions are dispersed and represent moments in the evolution of networks where a distribution of attributes in one text is dissimilar to the distribution in a later text to the extent that the revision can be considered as a re-write of the information contained in the first text. In this case, a path-dependent transition is indicated.

Critical transitions and the consequent path-dependencies are related to complex system dynamics and indicate to which extent a system is evolving following paths determined by previous states of the system. However, an evolving system can also be expected to "forget" parts of its history from the perspective of hindsight. Because we are using a literary model and not a behavioral one, we are able to address a dimension that may be latent to the authors who are entrained in the transition.

A critical transition is formally defined in terms of Shannon-type information. Shannon-type information is dimensionless and hence without meaning. It can be provided in this study with meaning as path-dependencies given the literary model. The meaning of documents thus identified can perhaps be validated in future research by interviews with experts (e.g., Campanario, 1993). Both main-path and critical transition results are combined below with the citation diagrams provided by *HistCite* in order to measure connectivity and evolutionary change along with prominence for each of the links and nodes.

**Results**

As noted, two sets of documents were retrieved from the ISI *Web of Science*:[4] the first, with "fullerene*" in their titles, contained 7,696 documents. The second set, based on selecting documents with "nanotube*" in their titles, contains 9,672 documents. The 30 most frequently cited documents in each of the sets and their internal relations were identified and illustrated with *HistCite* (Figures 3 and 4).

The analysis was limited to 30 documents for three reasons: (1) because the most often cited documents can be considered as central to the evolution of further research (Griffith *et al.,* 1974); (2) by selecting only the most cited papers we avoid loops in the citation network formed by all the retrieved documents; and (3) the limited sets enable to enhance visually the results obtained by *HistCite* without overcrowding the visualizations.

---

[4] All downloads were done between the 8th and the 17 of June of 2006. Documents from 1987 until 2005 were downloaded.

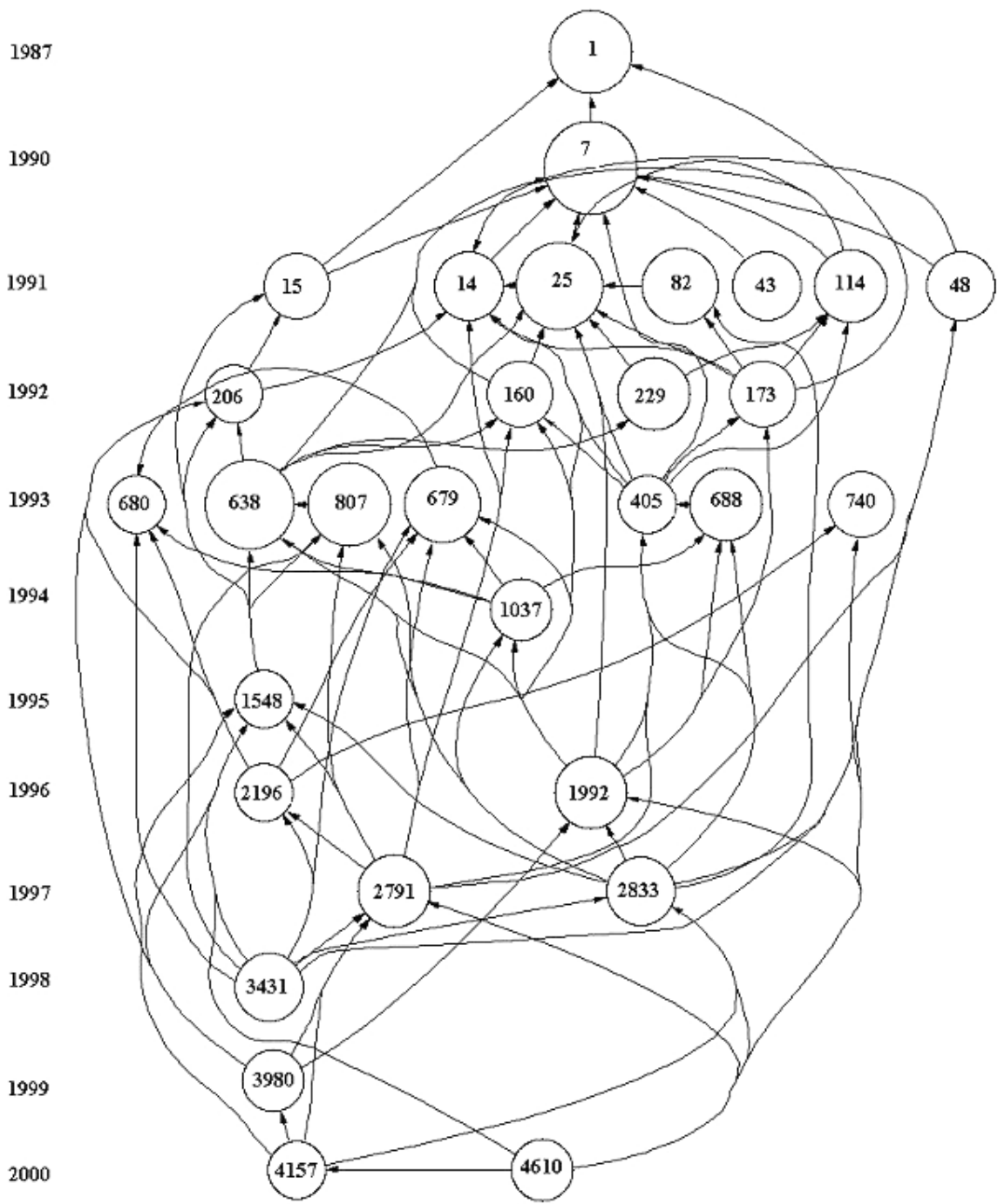

**Figure 3.** Thirty most highly-cited documents among 7,696 documents with "fullerene*" in their titles as generated by *HistCite*™.

| RANK | FIRST AUTHOR | YEAR | JOURNAL | CITATIONS |
|---|---|---|---|---|
| 1 | KROTO HW | 1987 | NATURE | 252 |
| 7 | TAYLOR R | 1990 | J CHEM SOC CHEM COMM | 303 |
| 14 | ALLEMAND PM | 1991 | J AMER CHEM SOC | 173 |
| 15 | HARE JP | 1991 | CHEM PHYS LETT | 163 |
| 25 | DIEDERICH F | 1991 | SCIENCE | 271 |
| 43 | HOWARD JB | 1991 | NATURE | 166 |
| 48 | ALLEMAND PM | 1991 | SCIENCE | 180 |
| 82 | CHAI Y | 1991 | J PHYS CHEM | 212 |
| 114 | DIEDERICH F | 1991 | SCIENCE | 182 |
| 160 | CREEGAN KM | 1992 | J AMER CHEM SOC | 162 |
| 173 | DIEDERICH F | 1992 | ACCOUNT CHEM RES | 149 |
| 206 | ANDERSSON T | 1992 | J CHEM SOC CHEM COMM | 119 |
| 229 | KIKUCHI K | 1992 | NATURE | 195 |
| 405 | ISAACS L | 1993 | HELV CHIM ACTA | 127 |
| 638 | TAYLOR R | 1993 | NATURE | 291 |
| 679 | FRIEDMAN SH | 1993 | J AMER CHEM SOC | 199 |
| 680 | SIJBESMA R | 1993 | J AMER CHEM SOC | 124 |
| 688 | BINGEL C | 1993 | CHEM BER-RECL | 182 |
| 740 | TOKUYAMA H | 1993 | J AMER CHEM SOC | 161 |
| 807 | MAGGINI M | 1993 | J AMER CHEM SOC | 238 |
| 1037 | HIRSCH A | 1994 | ANGEW CHEM INT ED | 147 |
| 1548 | WILLIAMS RM | 1995 | J AMER CHEM SOC | 130 |
| 1992 | DIEDERICH F | 1996 | SCIENCE | 186 |
| 2196 | JENSEN AW | 1996 | BIOORGAN MED CHEM | 131 |
| 2791 | IMAHORI H | 1997 | ADVAN MATER | 188 |
| 2833 | PRATO M | 1997 | J MATER CHEM | 167 |
| 3431 | PRATO M | 1998 | ACCOUNT CHEM RES | 171 |
| 3980 | DIEDERICH F | 1999 | CHEM SOC REV | 141 |
| 4157 | GULDI DM | 2000 | CHEM COMMUN | 121 |
| 4610 | GULDI DM | 2000 | ACCOUNT CHEM RES | 138 |

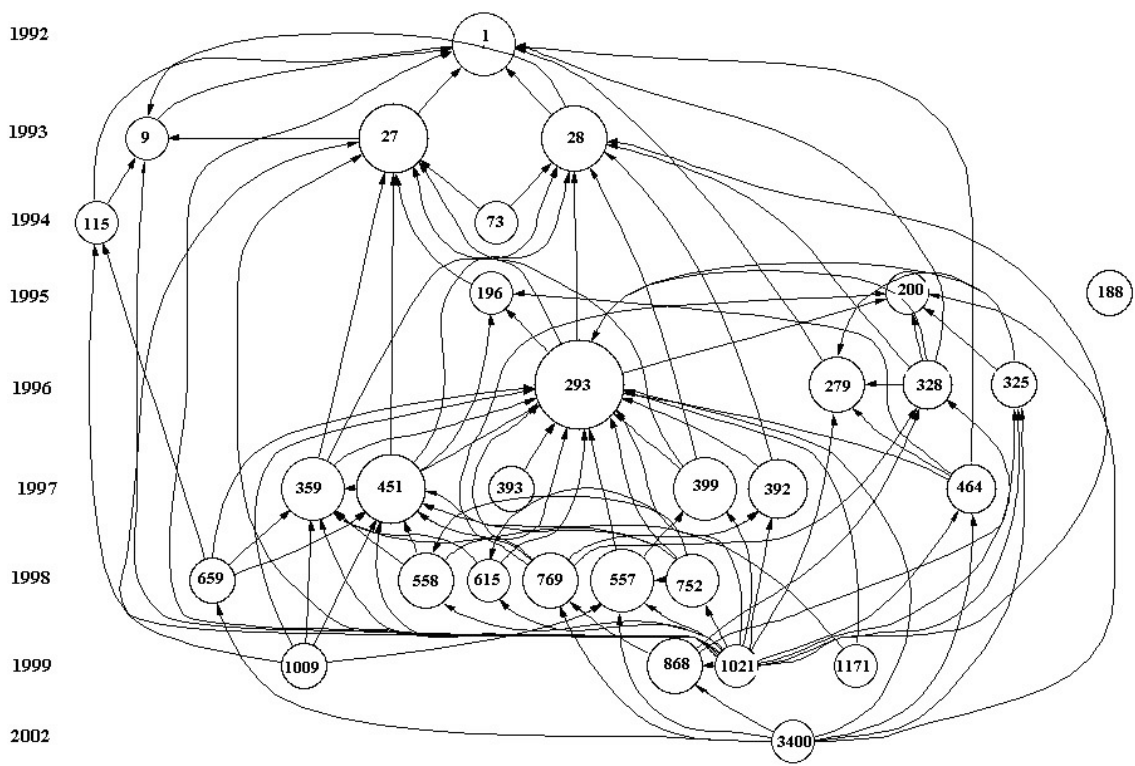

**Figure 4.** Thirty most highly-cited documents among 9,672 documents with "nanotube*" in their titles as generated by *HistCite*™.

| RANK | FIRST AUTHOR | YEAR | JOURNAL | CITATIONS |
|---|---|---|---|---|
| 1 | EBBESEN TW | 1992 | NATURE | 628 |
| 9 | AJAYAN PM | 1993 | NATURE | 298 |
| 27 | IIJIMA S | 1993 | NATURE | 760 |
| 28 | BETHUNE DS | 1993 | NATURE | 645 |
| 73 | BLASE X | 1994 | PHYS REV LETT | 275 |
| 115 | TSANG SC | 1994 | NATURE | 291 |
| 188 | CHOPRA NG | 1995 | SCIENCE | 325 |
| 196 | GUO T | 1995 | CHEM PHYS LETT | 286 |
| 200 | RINZLER AG | 1995 | SCIENCE | 287 |
| 279 | TREACY MMJ | 1996 | NATURE | 496 |
| 293 | THESS A | 1996 | SCIENCE | 1199 |
| 325 | DAI HJ | 1996 | NATURE | 334 |
| 328 | LI WZ | 1996 | SCIENCE | 388 |
| 359 | RAO AM | 1997 | SCIENCE | 588 |
| 392 | DILLON AC | 1997 | NATURE | 529 |
| 393 | BOCKRATH M | 1997 | SCIENCE | 335 |
| 399 | TANS SJ | 1997 | NATURE | 577 |
| 451 | JOURNET C | 1997 | NATURE | 718 |
| 464 | WONG EW | 1997 | SCIENCE | 387 |
| 557 | WILDOER JWG | 1998 | NATURE | 628 |
| 558 | ODOM TW | 1998 | NATURE | 512 |
| 615 | BANDOW S | 1998 | PHYS REV LETT | 278 |

| | | | | | |
|---|---|---|---|---|---|
| 659 | RINZLER AG | 1998 | APPL PHYS A | 319 |
| 752 | CHEN J | 1998 | SCIENCE | 434 |
| 769 | REN ZF | 1998 | SCIENCE | 484 |
| 868 | FAN SS | 1999 | SCIENCE | 485 |
| 1009 | KATAURA H | 1999 | SYNTHET METAL | 313 |
| 1021 | AJAYAN PM | 1999 | CHEM REV | 273 |
| 1171 | NIKOLAEV P | 1999 | CHEM PHYS LETT | 303 |
| 3400 | BAUGHMAN RH | 2002 | SCIENCE | 309 |

The *HistCite* algorithm is based on total cites of documents which can be expected to give a disadvantage to relatively recent papers because citations accumulate over time. Indeed, the distribution of the most relevant papers for the fullerene set of documents seems to indicate that the most highly cited papers in this field were written at its beginning (Figure 3). However, for the case of the nanotube documents (Figure 4), the distribution suggests a higher probability of becoming highly cited shortly after publication. Figure 3 illustrates the relevance of older documents in a vertical hierarchy-type of citation pattern. Notwithstanding the accumulative effects of citations to older documents in *HistCite,* Figure 4 shows a more horizontal distribution as well as a greater relevance of later documents. The nanotubes-field suggests sensitivity to change with the passing of time in the *HistCite* output.

Figure 5 depicts the *main path* of the 30 most often cited documents in the fullerene set as visualized by using Pajek. This result is used in Figures 6 to enrich the *HistCite* output; Figure 7 uses analogously the main-path analysis for the 30 most frequently cited documents in the nanotubes set.[5]

---

[5] When using the Search Path Node Pair instead of the Search Path Link Count, the citations relations in the early years were less overlapping.

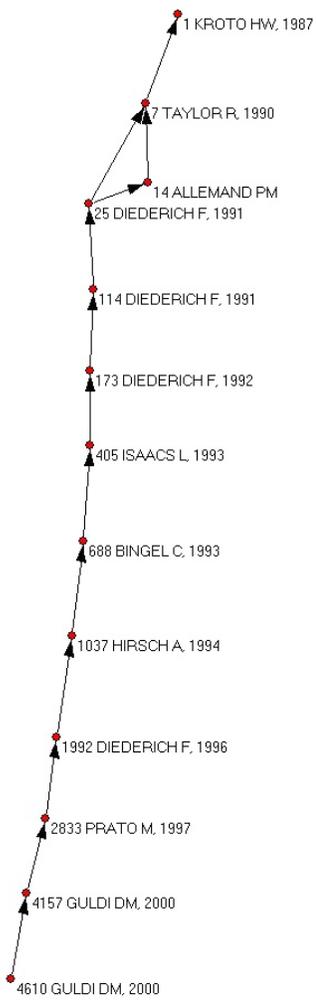

**Figure 5.** Main path for 30 most often cited documents in the set of "fullerene*"

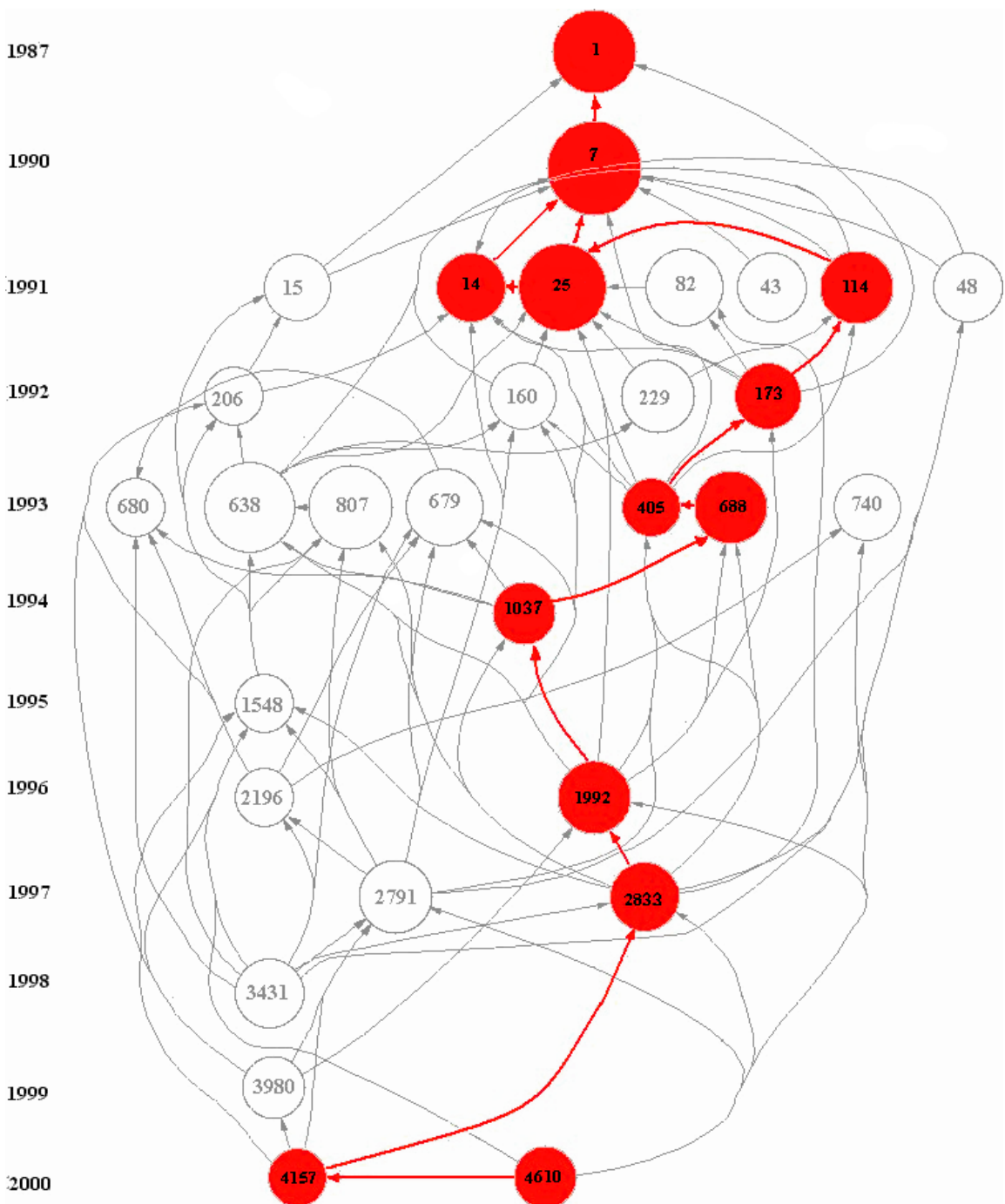

**Figure 6.** Main path for 30 most often cited documents in the set of "fullerene*" in *HistCite* output.

The main path for the set of fullerene documents (Figure 6) depicts a chronologically very stable structure, where the first and last paper are linked through the sequence of connecting nodes and the alternative paths dissolve after four years of publication. Thirteen documents, not necessarily the most highly cited ones, form the backbone of this network.

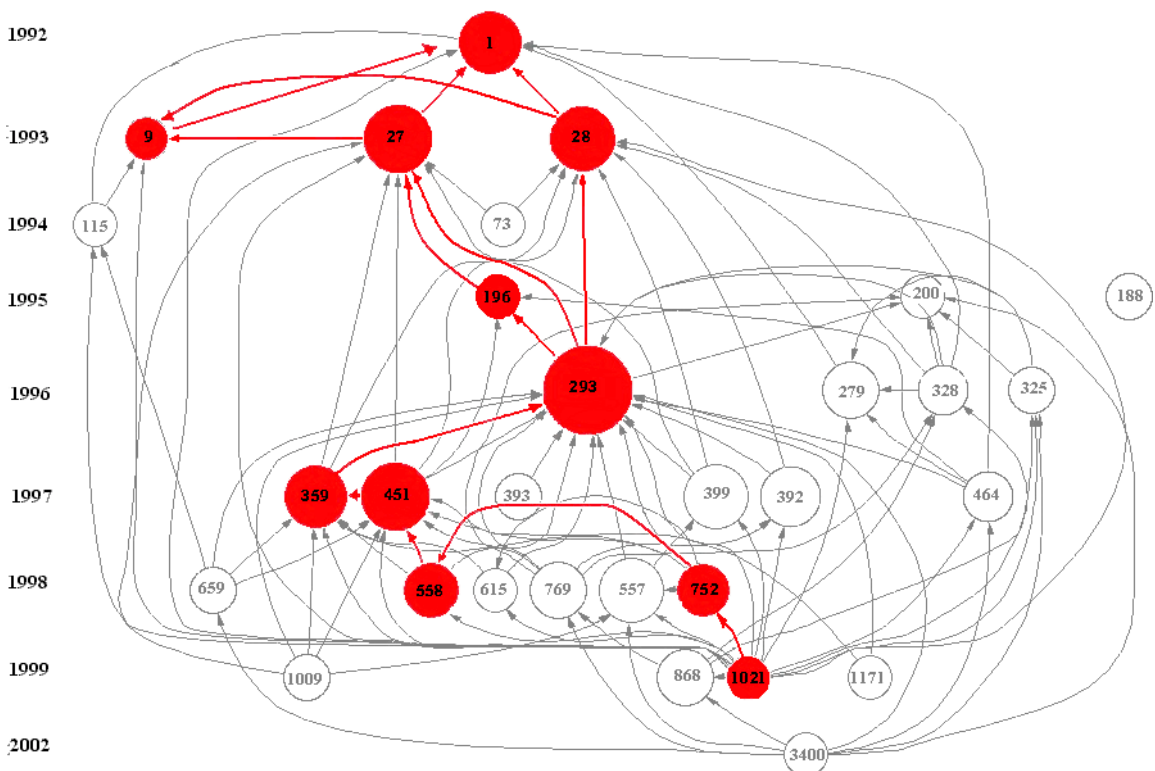

**Figure 7.** Main-path for 30 most often cited documents in the set of "nanotube*" in *HistCite* output.

For the set of nanotubes (Figure 7), the output is not as chronologically organized as for fullerenes and the main-path is less vertically structured. The largest nodes tend to be highlighted in the main path. Remember that the inclusion of a document in the main-path depends on the position of the paper among the 30 most highly cited documents, whereas the size of the node is determined (in *HistCite*) by the number of citations a document received in the full set. For the case of nanotubes, alternative main paths between nodes were found for the first six (of eleven) documents on the main path.

The critical transitions in terms of the distributions of cited references can be expected to indicate path-dependent transitions where an intermediate document holds more similarities with the distribution of the cited references in the later document. In the case of the most frequently cited documents of the set of fullerenes, the distributions were built using the 263 documents referenced more than once (among 1,034 references), and for the case of nanotubes, using the 91 references that occurred more than once (among 445 in total). As noted above, the Kullback-Leibler divergences among the distributions of cited references in each two texts were considered.

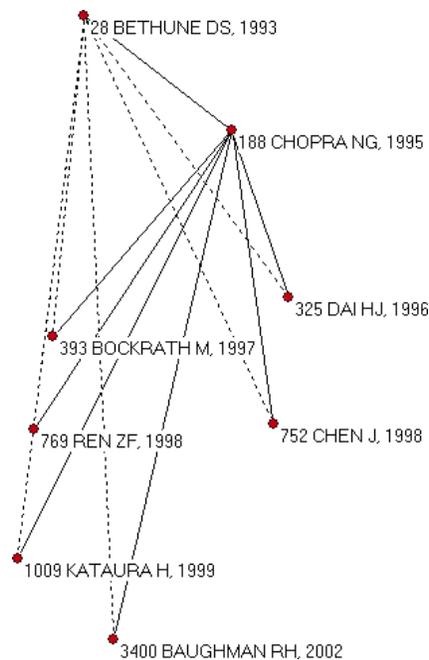

**Figure 8.** Path-dependent transitions in the distributions of cited references for the 30 most cited documents in the set of "nanotube*"

Figure 8 illustrates the results for the thirty most highly cited documents in the "nanotube*" set. The seven documents in the lower part of the picture are *a posteriori* to the paper by Bethune (1993). These cases suggest that an intermediate text (that is, Chopra, 1995) functions as a better predictor of the distribution of the cited references overwriting the distribution used in Bethune (1993). A closer look at the distributions reveals that all the documents in Figure 8 share four references used in Chopra (1995) as well as in Bethune (1993). However, in Bethunde (1993) nine other references were used that were no longer used by any of the *a posteriori* documents (cf. Price, 1965).

By using the citing direction for the analysis, one looks against the arrow of time, that is, from the perspective of hindsight. We have argued above that this visualization reflects the codification process involved in the positioning of published papers in bodies of knowledge already established: citations operate in this case as a retention mechanism (while in the forward direction they can be considered as a diffusion mechanism). History is continuously re-written as new papers appear providing references to the previous literature from the perspective of hindsight (*a posteriori*). The critical transitions represent significant events in the evolution of citation networks. At these moments, the evolutionary dynamics of the system changed in terms of the communication of cited references.

Note that not every critical transition depicted in Figure 8 represents also a citation relation in the *HistCite* diagrams. In fact, the intermediate document (188) was the only isolated node in Figure 4. In Figure 9 critical transitions in citation relations are penciled in as dotted black lines for the set of most highly cited documents in the domain of fullerenes. The critical revisions of the predictions are highlighted as documents 48 (1991), 679 (1993), 680 (1993), and 807 (1993)

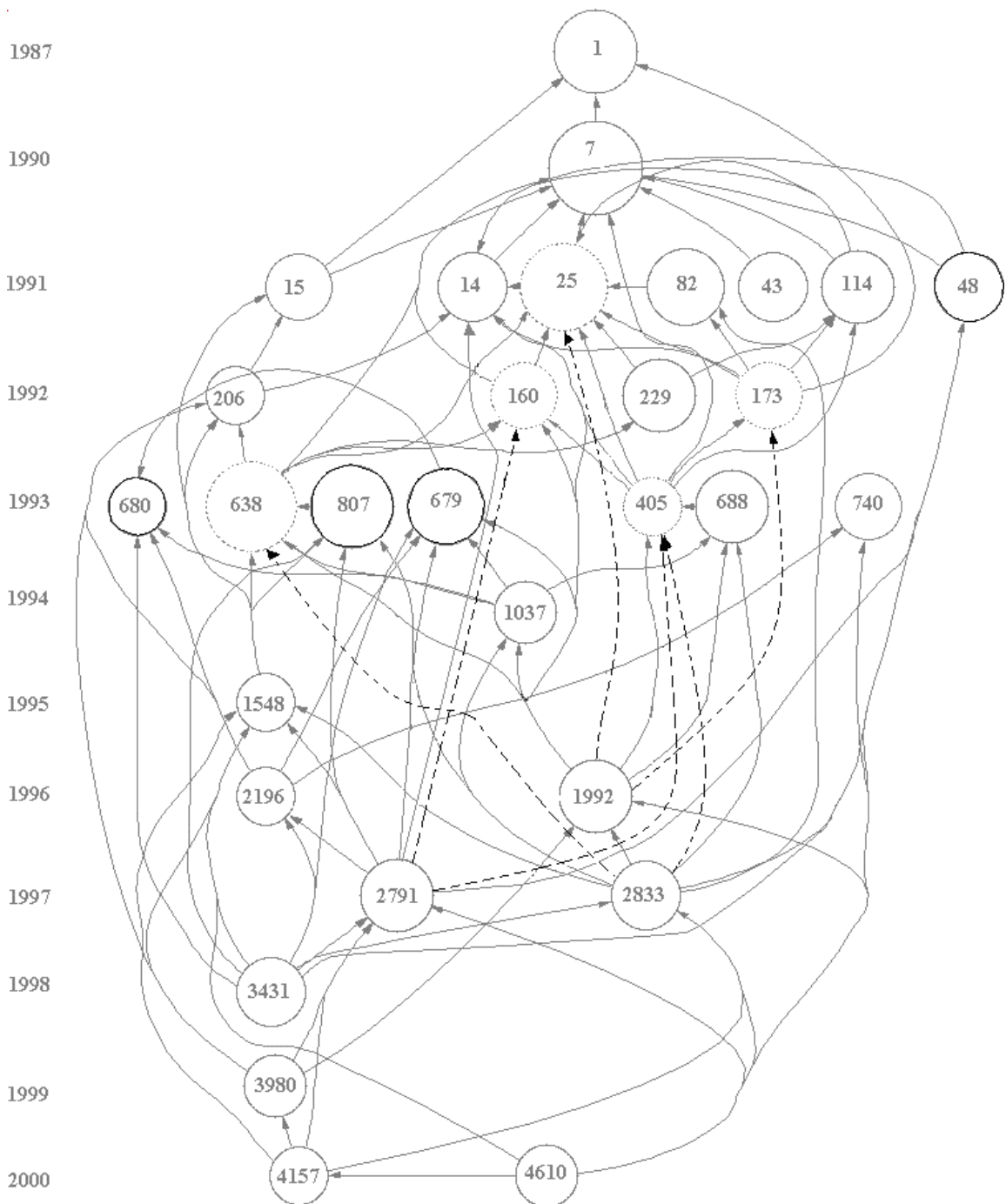

**Figure 9.** Critical transitions in the cited references distributions for the 30 most cited documents in the set of "fullerene*"

In the set of fullerenes, six links between nodes are dotted because an intermediate document (dark black) improved, for a citing text, the prediction of the distribution of the cited references in a cited text. In every case, the distribution of the cited references unravels over a period of five years; the intermediate document is chronologically closer to the first published (*a priori*) document than to the later one (*a posteriori*). Node 25 is

overwritten by 48, node 173 by 807, 160 by 680, node 638 by 679, and node 405 is overwritten by 680 and 679.

Because we are looking at similarities among the distributions of the cited references, the critical transitions indicate that it is no longer necessary to reach backwards behind the intermediate document. Documents published in 1991 and 1992 could be "forgotten" in 1996 because documents published in an in-between year (1993) had more similarities with the distribution of the cited references of documents published in 1996 and 1997. The relevant cognitive history has thus changed and path-dependency on the transition was generated.

This account hitherto represents a process of analyzing change as history evolves: looking at the publication of each document as an event that builds on previous publications, citations are analyzed from the perspective of the later published document. In order to understand scientific development and communication as a continuous process along the arrow of time, however, one needs to shift the perspective to how communication unfolds from the first published document.

**"Citing" *versus* "cited"**
By transposing the citation matrix obtained from *HistCite*, citations can also be analyzed in terms of "is cited by" relations that reflect the diffusion of ideas in the network of citations; this will yield results that differ from the study of the "citing" relations. While the "citing" relation indicates a codification process which makes the development of science reflexive on its own history (Garfield *et al.*, 1964), each citation can also be considered as a flow of ideas from cited documents to citing ones (Carley *et al.*, 1993). The "cited" relations reflect the diffusion of ideas from one paper into later ones.

Transposing the citation matrix and depicting the main stream of research as a flow of ideas from one paper to the next results in Figures 10 and 11. The process of scientific development is now coupled with the diffusion of early documents and their resonance in subsequent texts.

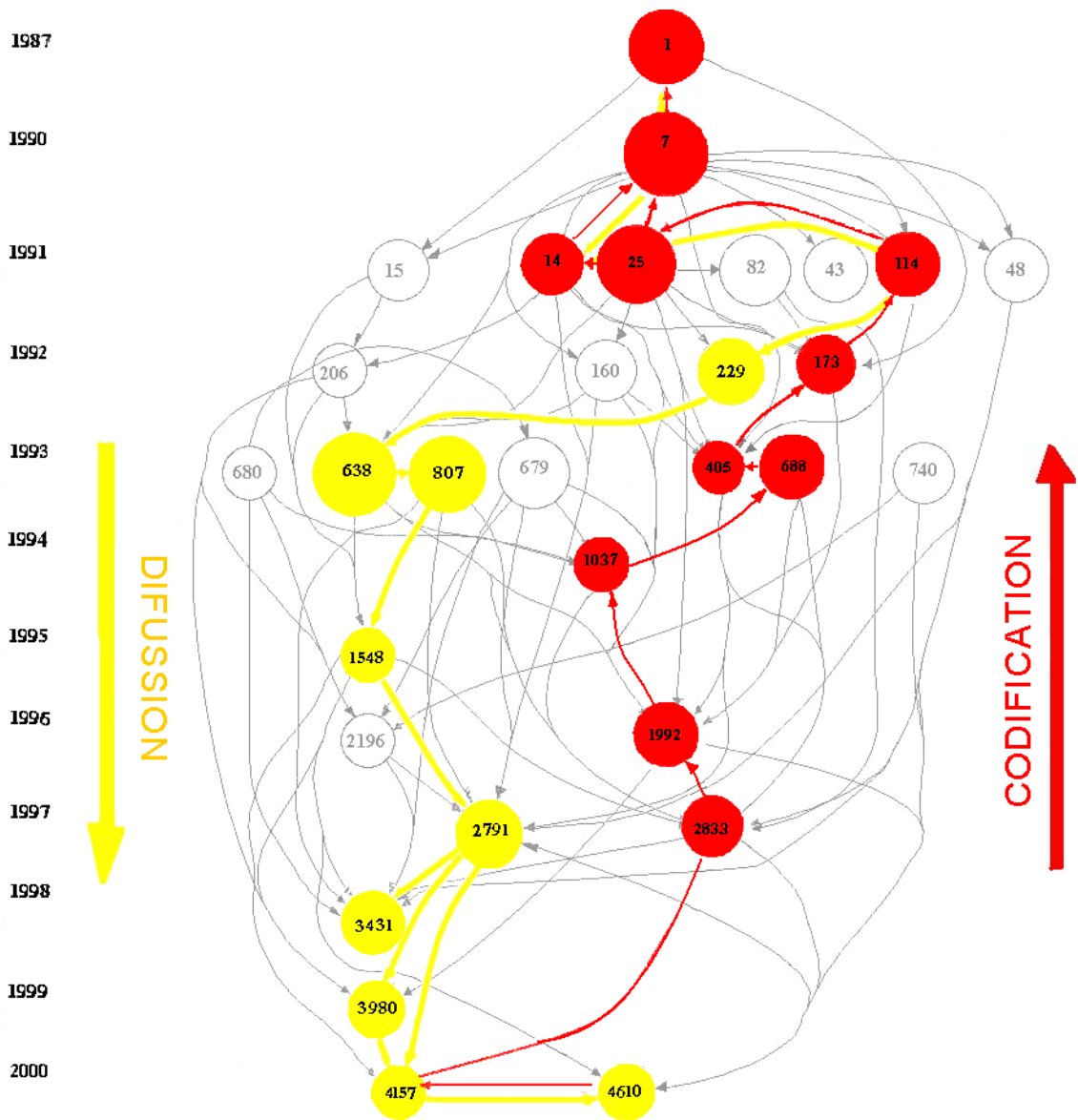

**Figure 10.** Most influential papers in the diffusion of research in fullerenes

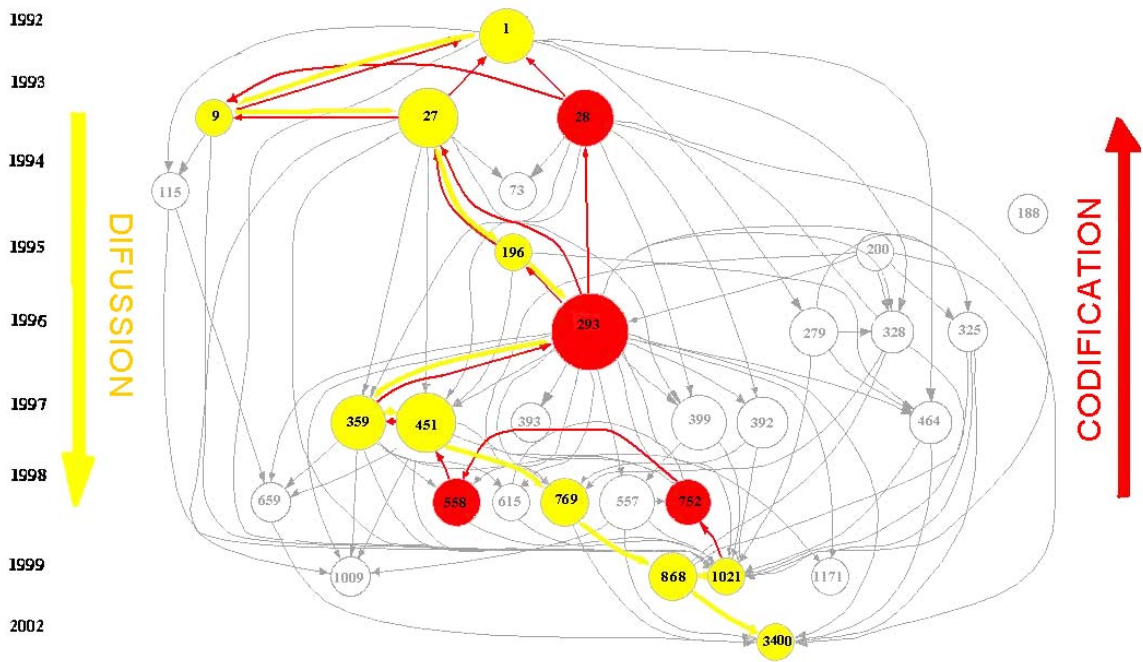

**Figure 11.** Most influential papers in the diffusion of research in nanotubes

Although many documents remain from the main paths of Figures 6 and 7 (nodes connected with darker arrows), others are no longer on the main path when the citation matrix is transposed. The Search Path Link Count is used again for depicting the new main path (lighter). In the previous figures, the main paths illustrated how earlier documents were codified in later ones so that central documents in the codification of the field were highlighted. Transposing the citation matrix, the main-path analysis selects the sequence of documents that illustrate the development of the research field by acting as the most influential documents over time (Carley *et al*., 1993). For the case of fullerenes (Figure 10), the change in the diffusion and codification process is noticeable from 1993 onwards, as new documents are included in the main path. This coincides with the moment when the rate of publications about nanotubes began to increase exponentially (Figure 1).

The transposed citation matrix can also be analyzed in terms of the expected information values that the information in the message—in terms of citation distributions—has changed during the diffusion process. For this, the distribution of the documents citing the 30 documents was considered to reflect the diffusion of each of these documents at the respective field level. For the set of "fullerenes," 1,281 documents cited more than one of the 30 most cited documents; for "nanotubes," this number was 3,224 documents. For the case of "fullerenes," a critical transition in terms of the distribution of documents citing the most cited documents was *never* the case. The fifteen critical transitions for the case of the distribution of the citing documents in the set of "nanotubes" are penciled into Figure 12.

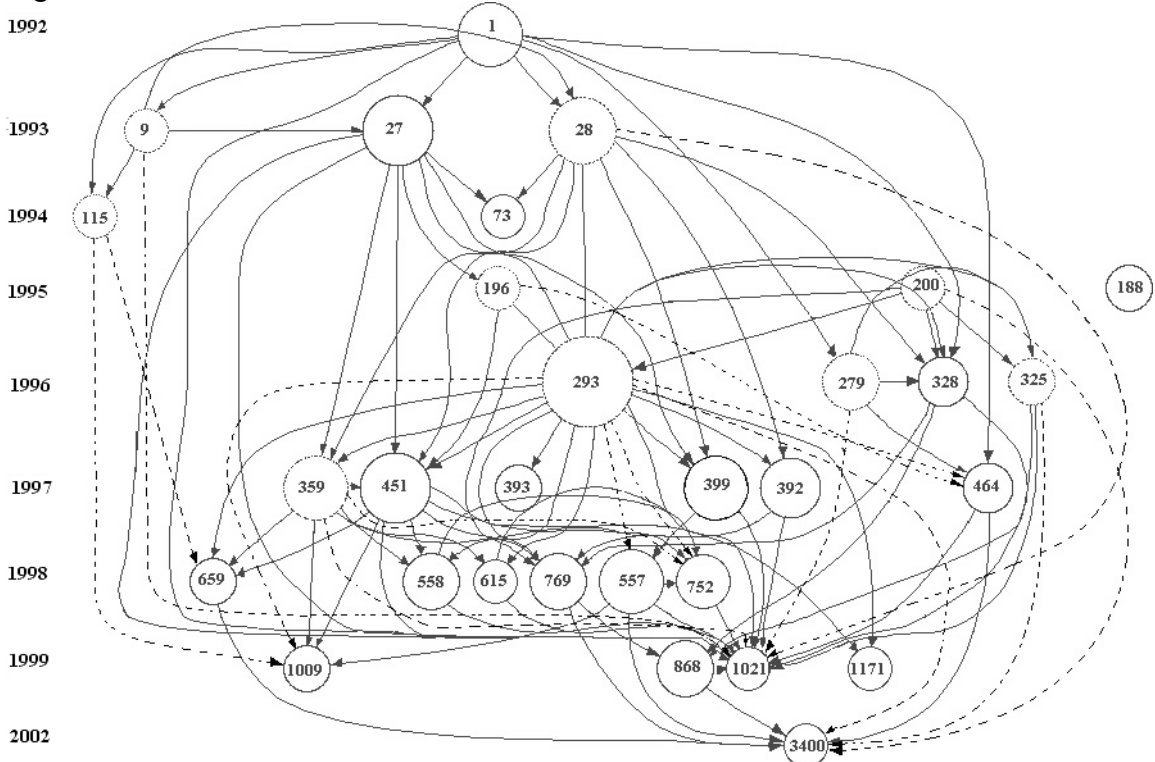

**Figure 12.** Critical transitions in the distributions of documents citing the 30 most cited documents in the set of "nanotube*"

The intermediate document acting as a better predictor of the distribution of the cited references is, in all cases, node 399 (highlighted) which can be an indicator of a very strong path dependency in 1997. Node 399 represents the paper by S. J. Tans *et al.* entitled "Individual single-wall carbon nanotubes as quantum wires," which appeared in *Nature* **386**, 474 - 477 (1997). Note that this paper is less highly cited than some of the other nodes (e.g., node 293).

The field of nanotubes emerged more recently than research in fullerenes, and has become a more dynamic topic of research, as illustrated by the growth in the publications rates in Figure 1. The citations to the 30 most highly cited documents in this network do not show signs of stabilization or accumulation. Dominant patterns have not emerged and sedimented as in the case of fullerenes, where the distribution of citations in citing documents to the 30 most cited documents seems to be shifting more gradually. For nanotubes, the emergence of similarities in the citations obtained by the 30 most highly cited documents suggests the formation of a solid intellectual base, but the citation patterns are still changing. This finding confirms our previous impression of stability in the fullerene historiogram and sensitivity to change in the dynamics of nanotubes. For research in nanotubes critical transitions in the diffusion process suggest evolutionary changes including important path-dependencies, but for the case of fullerenes the history of the whole field seems to remain relevant.

**Conclusions**

*HistCite*™ is a powerful tool that enabled us, among other things, to identify the most frequently cited works in any set of documents, and to build an inner-citation-matrix among them. The algorithm used to identify the most relevant documents is based on the frequency of citations over a period of time. *HistCite* visualizes the most relevant papers and the relations among them and can, therefore, be helpful for the historical reconstruction of scientific developments (Garfield *et al*., 2002, 2003a, 2003b).

Main-path algorithms (available in Pajek) identify which papers are most relevant in the overall flow of citations. The algorithm used in *HistCite* bases its estimations on frequencies. Main-path algorithms give priority to highly cited papers that have a considerable number of references as well. Papers that belong to the main path have been associated to thematic or methodological transitions in the development of a topic (Carley *et al.*, 1993). Documents highlighted by the main path can be considered as significant for writing the history of science and therefore also for researchers in the thus reconstructed field (Hummon & Doreian, 1989). The historian may have to argue why other paths were more important than the one visualized algorithmically.

Critical transitions provide indications of path dependency in the development of knowledge from an evolutionary perspective; the measurement of critical transitions can be considered as an operationalization of the concept of the evolutionary notion of a path dependency. It is expected that as time passes, some patterns will emerge in terms of distributions of attributes among texts, and some patterns will be forgotten as new texts continuously re-write the historical development of scientific fields.

The literary model can be used to simplify the process of scientific discovery and communication through the network of citation relations between scientific texts. The perspective of the scientist's activities is shifted from the micro-level of individual actions to the macro-structure resulting from the aggregate of those activities. *HistCite* allows us to operationalize the literary model for any defined set of documents by identifying the citation relations among them and illustrating the historical reconstruction of the set. Because *Histcite*'s output provides a chronologically ordered citation network, algorithms from social network analysis can be applied to further characterize the documents belonging to the set.

Main-path analysis allowed us to highlight which documents are central to the development of a specific topic either as codifiers of previously achieved ideas or as having a major influence on the development of the topic. The analysis of path dependencies provided further information about the transitional dynamics at research fronts. Research fronts develop by generating variation at specific moments of time. Main-path analysis focuses on the stabilization of structures over time. Critical transitions and path-dependencies can be expected to occur in parts of the current structure which are not yet part of the main path.

Our expectation was therefore that documents on the main path would not be "overwritten" by documents off this main path. We expected documents on the main path to overwrite documents outside the main path, but never to be overwritten because belonging to the main path summarizes the main findings in a field of research (Hummon & Doreian, 1989; Price, 1965). However, this was not the case. For example, node 405 of the fullerenes set (Figure 9) was part of the main path, but the information it provided with reference to nodes 2791 and 2833 was improved by later documents that themselves were not part of the main path.

The exercise of adding value to the links representing a citation relation was made in this study for two sets of documents retrieved from the ISI *Web of Science*: those with "fullerene*" in their title words and those with "nanotube*". Our concern with fullerenes and nanotubes started some time ago to illustrate changes in the dynamic construction of scientific knowledge. A scientific breakthrough can be characterized by an initially very dynamic moment which is followed by the emergence of new institutions as the topic gets attention on research agendas. As time passes, new hot topics emerge; some of them represent a bifurcation of the original scientific breakthrough (as seems the case for nanotubes), while others can be expected to stabilize.

In future research, we will propose to measure the self-organization of a topic of research as a reduction of uncertainty which may increase over time (Leydesdorff & Fritsch, 2006). In this paper we have focused on a citation network that chronologically links a set of published documents. In this way we could look at codification and diffusion as attributes of the documents belonging to the network. In a next paper, we will examine how the distribution of citations and title words co-evolve, and under what conditions this co-evolution leads to a reduction of uncertainty.

**Acknowledgements**
We are grateful to Dr. Eugene Garfield for providing us with a (beta-)version of HistCite™." We also thank the comments and suggestions made by anonymous referees.

return